\documentclass[doublecol]{epl2}
\title{Coevolutionary success-driven multigames}
\shorttitle{Coevolutionary success-driven multigames}
\author{Attila Szolnoki\inst{1} \and Matja{\v z} Perc\inst{2,3,4}}
\shortauthor{Szolnoki \& Perc}

\institute{\inst{1}Institute of Technical Physics and Materials Science, Research Centre for Natural Sciences, Hungarian Academy of Sciences, P.O. Box 49, H-1525 Budapest, Hungary\\
\inst{2}Faculty of Natural Sciences and Mathematics, University of Maribor, Koro{\v s}ka  cesta 160, SI-2000 Maribor, Slovenia\\
\inst{3}Department of Physics, Faculty of Science, King Abdulaziz University, Jeddah, Saudi Arabia\\
\inst{4}CAMTP -- Center for Applied Mathematics and Theoretical Physics, University of Maribor, Krekova 2, SI-2000 Maribor, Slovenia}


\abstract{Wealthy individuals may be less tempted to defect than those with comparatively low payoffs. To take this into consideration, we introduce coevolutionary success-driven multigames in structured populations. While the core game is always the weak prisoner's dilemma, players whose payoffs from the previous round exceed a threshold adopt only a minimally low temptation to defect in the next round. Along with the strategies, the perceived strength of the social dilemma thus coevolves with the success of each individual player. We show that the lower the threshold for using the small temptation to defect, the more the evolution of cooperation is promoted. Importantly, the promotion of cooperation is not simply due to a lower average temptation to defect, but rather due to a dynamically reversed direction of invasion along the interfaces that separate cooperators and defectors on regular networks. Conversely, on irregular networks, in the absence of clear invasion fronts, the promotion of cooperation is due to intermediate-degree players. At sufficiently low threshold values, these players accelerate the erosion of defectors and significantly shorten the fixation time towards more cooperative stationary states. Coevolutionary multigames could thus be the new frontier for the swift resolution of social dilemmas.}

\begin{document}

\maketitle

Evolutionary multigames \cite{hashimoto_jtb06, hashimoto_jtb14} or mixed games \cite{wardil_csf13} are one of the more recent and very promising developments in evolutionary game theory \cite{maynard_82, weibull_95, hofbauer_98, mestertong_01, nowak_06}. Inspired by the early research on dynamic stability in symmetric games with two decisions \cite{cressman_igtr00},
evolutionary multigames enable the usage of different payoff matrices in the realm of the same competitive process. This might be particularly useful in the realm of social dilemmas, where the pursuit of short-term individual benefits may quickly result in the erosion of cooperation, and ultimately also in the ``tragedy of the commons'' \cite{hardin_g_s68}. However, when individuals are torn between what is best for them and what is best for the society, one should not overlook the importance of personal success and status. To illustrate the point, imagine two individuals wanting to use public transport. One is wealthy and the other is poor. Under normal circumstances, the wealthy individual is less likely to defect by using the services without paying for the ticket. Put differently, the temptation to defect is certainly higher for the poor individual. Thus, when we face a dilemma, we are likely to perceive differently what we might gain by choosing to defect, and evolutionary multigames provide the theoretical framework for properly addressing precisely such situations.

No social dilemma has received as much attention as the prisoner's dilemma game \cite{fudenberg_e86, nowak_n93, santos_prl05, wang_wx_pre08, santos_pnas06, fu_epjb07, tanimoto_pre07, gomez-gardenes_prl07, pacheco_ploscb09, fu_pre08b, wang_wx_c11,  antonioni_pone11, rong_pre10, fu_pre09, tanimoto_pre12, press_pnas12, hilbe_pnas13, fu_jtb10, yang_hx_njp14, szolnoki_pre14, santos_md_srep14, wu_js_pa14}, and this is why we also adopt it as the core game in this letter. Each instance of the game is contested by two players who have to decide simultaneously whether they want to cooperate or defect, and the dilemma is given by the fact that although mutual cooperation yields the highest collective payoff, a defector will do better if the opponent decides to cooperate. Due to the different sources of heterogeneity, however, the symmetry in payoff elements is not necessarily fulfilled \cite{wang_j_pre10b, hauser_jtb14}.
The key question that we aim to answer is, how the introduction of different payoff matrices that are used by individual players based on their success in previous rounds of the game will influence the cooperation level in the whole population? Although previous research has focused on situations where many different games are played simultaneously \cite{cressman_igtr00, hashimoto_jtb06, hashimoto_jtb14, wardil_csf13}, in the present work we consider identical games, but with different payoff values. Furthermore, we consider a coevolutionary setup \cite{perc_bs10}, where along with the strategies the perceived strength of the temptation to defect coevolves with the success of each individual player.

As we will show, on regular networks success-driven selection of the temptation to defect leads to dynamically reversed direction of invasion along the interfaces that separate cooperators and defectors, while on irregular networks the promotion of cooperation is due to intermediate-degree players that accelerate the erosion of defectors. This is significantly different from the expected outcome in well-mixed populations, where the stationary state can be deduced by the average values of the employed payoff elements. The study of coevolutionary multigames in structured populations thus highlights the usefulness of interdisciplinary research centered around statistical physics, which has recently revealed many new ways by means of which social dilemmas could be resolved in favor of cooperative behavior \cite{szabo_pr07, roca_plr09, schuster_jbp08, perc_bs10, santos_jtb12, perc_jrsi13, rand_tcs13, wu_t_pa14, van-doorn_jtb14}.

Formally, we study pairwise evolutionary games, where mutual cooperation yields the reward $R$, mutual defection leads to punishment $P$, and the mixed choice gives the cooperator the sucker's payoff $S$ and the defector the temptation $T$. Within this setup we have the weak prisoner's dilemma game if $T>R>P=S$. Without loss of generality we set $R=1$, $P=S=0$ and $T>1$. However, $T>1$ is a free parameter only for those players $x$ whose payoff from the previous round $\Pi_x^p$ is not larger than a threshold $\Theta$. If $\Pi_x^p > \Theta$, then player $x$ uses only a minimally low temptation to defect $T_1<T$ in the next round of the game. We will use $T_1=1.01$ as a fixed value, which still ensures a higher payoff for defectors and thus preserves the weak prisoner's dilemma ranking at all times.

We use a $L \times L$ square lattice with periodic boundary conditions as the simplest interaction network to describe a structured population, and we also consider scale-free networks of size $N$ as a somewhat more apt description of realistic social and technological networks \cite{barabasi_s99}. In what follows, we will specify the applied system size with the presented results, although we emphasize that our results are robust and valid in the large system size limit, when the latter sufficiently exceeds the typical correlation length of the emerging spatial patterns. To make the results on different networks comparable, we apply threshold values that are proportional to their average degree $<z>=z_a$, such that $\Theta=z_aH$ where $H$ denotes the scaling factor of the threshold.

Unless stated differently, for example to illustrate a specific invasion process as in Fig.~\ref{stripe}, we use random initial conditions such that cooperators $s_x=C$ and defectors $s_x=D$ are uniformly distributed across the network. We simulate the evolutionary process in accordance with the standard Monte Carlo simulation procedure comprising the following elementary steps. First, a randomly selected player $x$ acquires its payoff $\Pi_x$ by playing the game with all its neighbors. Next, player $x$ randomly chooses one neighbor $y$, who then also acquires its payoff $\Pi_y$ in the same way as previously player $x$. Lastly, player $x$ adopts the strategy $s_y$ from player $y$ with a probability determined by the Fermi function
\begin{equation}
\label{eq1}
W(s_y \to s_x)=\frac{1}{1+\exp[(\Pi_x-\Pi_y)/K]},
\end{equation}
where $K=0.1$ quantifies the uncertainty related to the strategy adoption process \cite{szabo_pr07}. The selected value ensures that better-performing players are readily followed by their neighbors, although adopting the strategy of a player that performs worse is not impossible either. This accounts for imperfect information, errors in the evaluation of the opponent, and similar unpredictable factors. Each full Monte Carlo step (MCS) gives a chance for every player to change its strategy once on average. All simulation results are obtained on networks with $N=10^4 - 2 \cdot 10^5$ players or more depending on the proximity to phase transition points, and the fraction of cooperators $f_C$ is determined in the stationary state after a sufficiently long relaxation extending up to $2 \cdot 10^5$ MCS. To further improve accuracy, the final results are averaged over $1000$ or more independent runs.

Before turning to the results, it is worth pointing out that the selection of the temptation for each particular player $x$ depends only on its payoff from the one previous round $\Pi_x^p$. It is possible to extend the ``memory'' of players and thus to consider payoffs from $M>1$ previous rounds, but the main conclusions remain the same as presented below. Notably, considering $M>1$ still does not involve traditional ``memory effects'' in evolutionary games because player $x$ does not record past strategies of its neighbors, but simply accumulates own payoffs from the $M$ previous rounds. We also note that we have explored structured populations other than those described by the square lattice and the scale-free network, and we have also found that the main results remain robust.

\begin{figure}
\begin{center}
\includegraphics[width = 8.3cm]{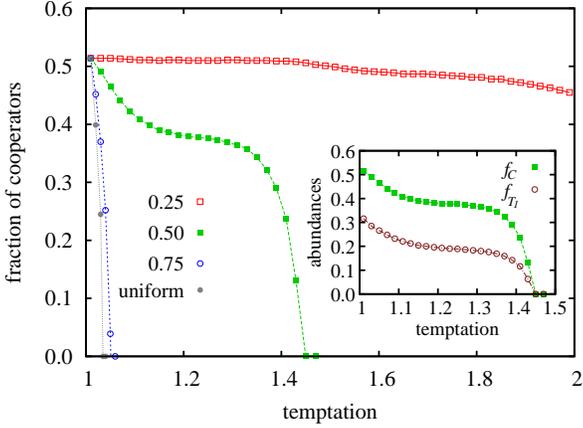}
\caption{\label{threshold} The fraction of cooperators $f_C$ in dependence on the temptation $T$, as obtained for different threshold scaling factors $H$ (see legend) on the square lattice. For comparison, we also show results obtained with the uniform (classical) model, where every player $x$ uses the same temptation $T$, regardless of its success in the previous round -- irrespectively of $\Pi_x^p$. The inset shows the fraction of all players using $T_1$ (satisfying $\Pi_x^p > \Theta$), along with the fraction of cooperators $f_C$, as obtained for $H=0.5$. We have used lattices with liner size $L=400$.}
\end{center}
\end{figure}

\begin{figure}
\begin{center}
\includegraphics[width = 8.3cm]{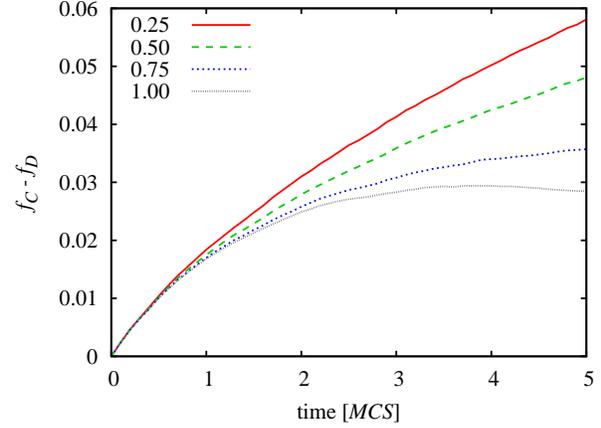}
\caption{\label{stripe} The competition between cooperators and defectors ($f_C-f_D$) along initially straight interfaces, as obtained for different threshold scaling factors $H$ (see legend) on the square lattice. We have used prepared initial states, such that cooperators and defectors were arranged in stripes of width $W=20$ on a square lattice with linear size $L=1000$. It can be observed that initially, when straight interfaces maximize the efficiency of network reciprocity, cooperators are more successful than defectors regardless of the value of $H$. As time goes on and the interfaces roughen, however, sufficiently small values of $H$ are able to reverse the direction of invasion and keep it in favor of cooperators. Note that for larger values of $H$ the cooperators start giving in to defectors, and eventually $f_C=0$ will be reached for $H=0.75$ and $H=1$ (see Fig.~\ref{threshold}). We have used the temptation $T=1.3$.}
\end{center}
\end{figure}

We begin by exploring how the fraction of cooperators $f_C$ varies in dependence on the temptation $T$ and the threshold scaling factor $H$. As shown in the main panel of Fig.~\ref{threshold}, the lower the value of $H$, the larger the value of $T$ at which cooperators are still able to survive. Naturally, at $H=0$ everybody would play the game with the minimal temptation $T_1=1.01$, while for larger values of $H$ players will increasingly often use the higher $T$ value. When the threshold $\Theta=z_aH$ exceeds the maximally attainable payoff for an individual player, we recover the uniform (classical) model, where every defector uses $T$ at all times.

If the game were contested in a well-mixed population, the reported promotion of cooperation in Fig.~\ref{threshold} would simply be the consequence of an overall lower average value of $T$ (see \cite{wardil_csf13} for details). In structured populations, however, this is not the case because here only a minority of players is using the smaller $T_1<T$ value. To demonstrate this, we show in the inset of Fig.~\ref{threshold} the fraction of those players who use $T_1$ instead of the higher $T$ value when calculating their payoffs, as obtained for $H=0.5$. For convenience, we also plot $f_C$ in dependence on $T$ at the same $H$ value. Interestingly, the fraction of players using $T_1$ is getting smaller as $T$ increases, and it lastly drops to zero when the system terminates into the all-$D$ phase. Of course, the extinction of those using $T_1$ at that point is understandable since in the absence of cooperators there is no one to exploit, and hence everybody has payoff zero and fails to exceed $\Theta=z_aH$.

To understand the reported promotion of cooperation, instead of a ``static'' or equilibrium argument that works for well-mixed populations, in our case we have to look more carefully at the microscopic dynamics. We find that, in fact, the positive effect has a dynamical origin: when a defector collects a large payoff that satisfies $\Pi_p > z_aH$, then its next payoff will be lower due to the usage of the smaller $T_1$ value. Consequently, defectors fail to spread effectively, which ultimately gives rise to a higher stationary fraction of $C$ players. This effect can be demonstrated transparently by monitoring the movement of interfaces between competing strategies, where in the prepared initial state the strategies are arranged into stripes. The evolution of cooperation from such special initial conditions is plotted in Fig.~\ref{stripe} for different values of $H$. Initially, when the interfaces are straight, cooperators have an advantage regardless of $H$ (all curves rise) because they are supported by almost all their neighbors, while at the same time the neighboring defectors can collect a high payoff from only a single $C$ neighbor. In other words, network reciprocity is in full effect \cite{nowak_n92b}. But after a short time, when interfaces become irregular, the evolution depends sensitively on the value of $H$. While for $H=0.75$ and $H=1$ defectors start spreading into the territory of cooperators to eventually yield the $f_C=0$ phase, for sufficiently low values of $H$ we can observe a dynamically reversed direction of invasion along the interfaces that separate cooperators and defectors. Cooperators are therefore able to survive at adverse conditions that are beyond reach for network reciprocity alone.

\begin{figure}
\begin{center}
\includegraphics[width = 8.3cm]{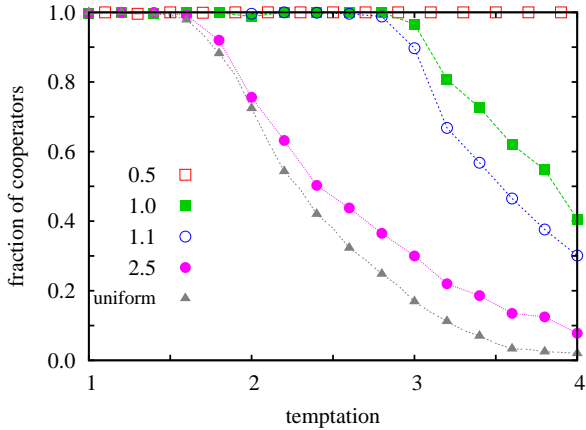}
\caption{\label{sf} The fraction of cooperators $f_C$ in dependence on the temptation $T$, as obtained for different threshold scaling factors $H$ (see legend) on the scale-free network. We have used networks with $N=10^4$ players, and the displayed results were averaged over $5000$ independent runs, so that the error bars are comparable with the size of the symbols.}
\end{center}
\end{figure}

On irregular networks, where clear invasion fronts are absent, one may expect slightly different behavior from what we have reported above for the square lattice (qualitatively identical result can be obtained also for other regular networks). To demonstrate this, we show in Fig.~\ref{sf} results obtained on the scale-free network. It can be observed that, like on the square lattice, the lower the value of $H$, the larger the value of $T$ at which cooperators are still abundant in the population. We note that, for clarity, the considered temptations in Fig.~\ref{sf} significantly exceed the traditional $T=2$ prisoner's dilemma limit. Moreover, because of the presence of hubs, we need a much higher value of $H$ to recover the uniform (classical) model, where everybody uses $T>T_1$ at all times. But this also means that we can allow more defectors to use the higher $T$ value (apply larger $H$ values) and still have a highly cooperative society.

\begin{figure}
\begin{center}
\includegraphics[width = 8.3cm]{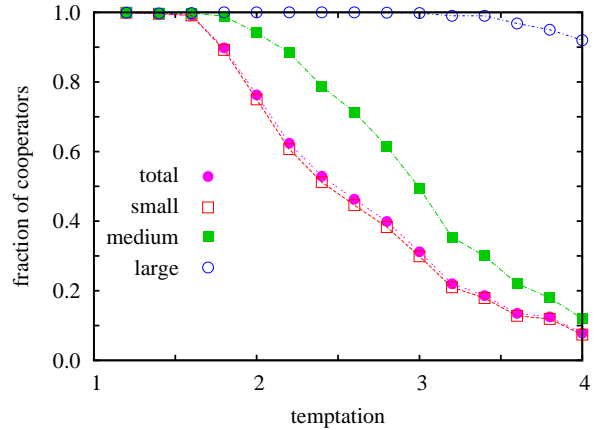}
\caption{\label{class} The fraction of nodes that are occupied by cooperators in dependence on the temptation $T$, as obtained for $H=2.5$ on the scale-free network. Nodes are classified into three categories, namely for having small, medium and large degree. For comparison, we also show the overall level of cooperation in the network.}
\end{center}
\end{figure}

This additional promotion of cooperation might be related to the heterogeneity of the interaction topology \cite{santos_jtb12}, where it is generally believed that the highly connected players rule the evolutionary process \cite{santos_prl05, santos_pnas06, gomez-gardenes_prl07}. In our case, the hubs can easily collect high payoffs, and therefore they will always apply the low $T_1$ value. This, however, weakens only the defectors that occupy the hubs, but not the cooperators.
But importantly, this statement is valid for almost any $H$ value, and hence it can not explain the strong dependence of the cooperation level on $H$ that we report in Fig.~\ref{sf}.
Our argument can be supported convincingly by comparing the cooperation level among players that have the same ``degree class''. We therefore divide the players into three categories, such that the degree span belonging to each group is equally large on the logarithmic scale. This allows us to determine the fraction of cooperators among players having small, medium and large degree. A representative comparison is plotted in Fig.~\ref{class}. While the cooperation level remains high among the hubs, they cannot influence the rest of the population, and accordingly, the average cooperation level may decrease significantly at high values of $T$. This behavior is significantly different from the previously uncovered mechanism that is responsible for augmented cooperation levels on heterogeneous networks when uniform payoff values were applied \cite{santos_prl05, szabo_pr07}. In the latter case, the asymmetric information flow where the hubs were always followed by small-degree players is essential. In our case, however, the crucial difference is due to the behavior of intermediate-degree players, who are responsible for spreading the successful strategy from hubs towards the abundant ``minor'' (weakly connected) players. If defectors could remain powerful among the middle class, then they could block the invasion of cooperation. But if intermediate-degree defectors are weakened due to the application of a sufficiently low value of $H$, then only cooperative hubs can ``spread'' their strategy successfully across the whole population.

\begin{figure}
\begin{center}
\includegraphics[width = 8.3cm]{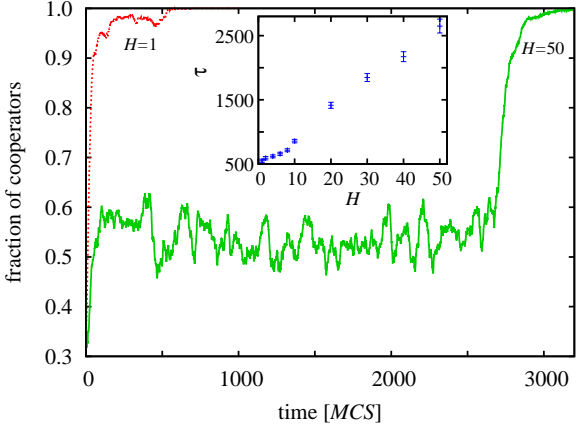}
\caption{\label{fix} Representative time evolution of the fraction of cooperators $f_C$, as obtained for $H=1$ (dotted red line) and $H=50$ (solid green line) at $T=1.5$ if starting from a random initial state on scale-free network. The inset shows the average fixation time $\tau$ in dependence on threshold scaling factors $H$. Supporting the results in the main panel, it can be observed that the fixation time needed to reach $f_C=1$ decreases significantly as $H$ decreases. We have used scale-free networks with $N=10^4$ players, and the displayed results were averaged over $5000$ independent runs.}
\end{center}
\end{figure}

To corroborate the above argumentation, we compare the evolutionary dynamics at different values of $H$, but when the final outcome is identical. Results presented in Fig.~\ref{sf} show that at $T=1.5$ the system always evolves to a cooperator-dominated state, independently of the $H$ value. However, results presented in Fig.~\ref{fix} demonstrate that the dynamics to reach the $f_C=1$ phase does depend sensitively on the applied threshold. At $H=50$ (solid green line), where due to the high threshold intermediate-degree defectors remain relatively strong, the resistance against cooperators persists much longer than at $H=1$. Ultimately, $f_C=1$ is always reached, but the fixation time at $H=1$, where middle class defectors are forced to use the minimally low $T_1<T$ value, is much shorter. Naturally, each individual fixation time can depend sensitively on the generated network and on the initial distribution of strategies. Therefore, to obtain conclusive results, we have averaged the fixation time over a large number of independent runs. As the inset of Fig.~\ref{fix} shows, the average fixation time indeed increases unambiguously with increasing values of $H$, and this despite of the fact that the behavior of the leading hubs remains unchanged.

\begin{figure}
\begin{center}
\includegraphics[width = 8.3cm]{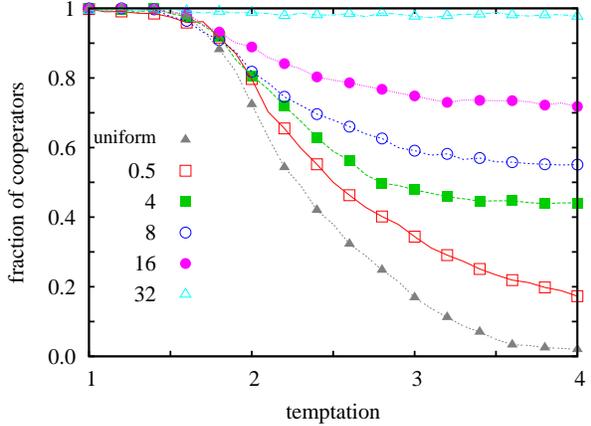}
\caption{\label{reverse} The fraction of cooperators $f_C$ in dependence on the temptation $T$, as obtained for different threshold scaling factors $H$ (see legend) on the scale-free network. Here the reversed coevolutionary rule is applied, such that ``wealthy'' players who fulfill $\Pi_x^p > \Theta$ use $T>T_1$, while the ``poor'' players use $T_1<T$. Evidently, the evolution of cooperation can be promoted just as well as with the original coevolutionary rule, for which results are presented in Fig.~\ref{sf}.}
\end{center}
\end{figure}

The important role of intermediate-degree players, or alternatively, only the second-order importance of hubs (which is an unexpected results by itself), can be verified effectively if we modify the microscopic dynamics so that ``wealthy'' players, those whose payoffs fulfill $\Pi_x^p > \Theta$, use the actual $T$ value while the unsuccessful players use the minimally low $T_1<T$. Although this is an obviously ``unfair'' rule where the cooperators are expected to struggle, it is perfectly suited for our purposes at this point. Moreover, because of the applied coevolutionary rule, cooperators actually still fare very well against the defectors. Indeed, the results obtained with the reversed coevolutionary rule presented in Fig.~\ref{reverse} are comparable to the results obtained with the original coevolutionary rule shown in Fig.~\ref{sf}. Naturally, in the reversed case, we recover the uniform (classical) results on the scale-free network only at the small threshold limit, when all players use the higher $T$ value at all times.

\begin{figure}
\begin{center}
\includegraphics[width = 8.3cm]{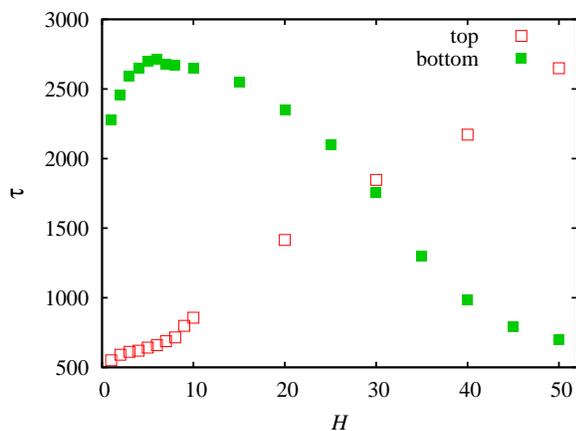}
\caption{\label{compare} The average fixation time $\tau$ in dependence on the threshold scaling factor $H$, as obtained for the original coevolutionary rule (denoted as ``top'') and for the reversed coevolutionary rule (denoted as ``bottom'') at $T=1.4$. We have used scale-free networks with $N=10^4$ players. The displayed results were averaged over $5000$ independent runs, so that the error bars are comparable with the size of the symbols.}
\end{center}
\end{figure}

To further demonstrate the decisive role of the middle class -- players with intermediate degree within the network -- we have again measured the fixation times at different threshold scaling factors $H$ when the final state is an all-$C$ phase. For the reversed coevolutionary rule, however, we have used a slightly smaller $T=1.4$ value than for the original coevolutionary rule (see Fig.~\ref{fix}). As the results presented in Fig.~\ref{compare} show, for the reversed coevolutionary rule too, there is again a significant difference in how fast the same final state is reached. Indeed, a fast transition to the pure $C$ phase can only be reached when intermediate-degree defectors are ``weakened'' by the necessity to use the lower threshold value.

Summarizing, we have shown that the introduction of coevolutionary multigames in structured populations reveals new ways for the resolution of social dilemmas. While this may be an expected consequence of the fact that players in structured populations have a limited interaction range, more importantly, our model also illustrates that the stationary state in coevolutionary multigames can not be deduced simply by the average values of the employed payoff elements. On regular networks, such as the square lattice as the simplest approximation of a structured population, cooperation benefits from dynamical reversals of the spreading direction of interfaces that separate cooperators and defectors. We have demonstrated this by studying the evolution of cooperation with prepared initial conditions, which have revealed accelerated and persistent interface expansion towards the region of defectors. However, on irregular networks, such as the scale-free network as a simple theoretical proxy for social and technological networks, cooperation benefits from the middle class players, which significantly accelerate the erosion of defectors, thereby shortening the fixation times towards highly cooperative states. Interestingly, we have shown that, even if successful players adopt higher temptation values and unsuccessful players adopt the minimally low temptation values, the evolution of cooperation is still promoted significantly. Thus, it appears that the details of the coevolutionary process are of secondary importance. What matters most is simply the possibility for individual players to express their differing perception of the situation, without much regard for the details of the perception itself.

We note that all the presented results are robust to variations of the interaction network, and can be observed also if the payoffs are accumulated longer from the past and if the social dilemma is different from the prisoner's dilemma, overall indicating a high degree of universality. We conclude that coevolutionary multigames might be the new frontier for the identification of new mechanisms that lead to the swift resolution of social dilemmas.

\acknowledgments This research was supported by the Hungarian National Research Fund (Grant K-101490), John Templeton Foundation (FQEB Grant \#RFP-12-22), TAMOP-4.2.2.A-11/1/KONV-2012-0051, and the Slovenian Research Agency (Grants J1-4055 and P5-0027).


\begin{thebibliography}{10}
\expandafter\ifx\csname url\endcsname\relax\def\url#1{\texttt{#1}}\fi

\bibitem{hashimoto_jtb06}
\Name{Hashimoto K.} \REVIEW{J. Theor. Biol.}{241}{2006}{669}.

\bibitem{hashimoto_jtb14}
\Name{Hashimoto K.} \REVIEW{J. Theor. Biol.}{345}{2014}{70}.

\bibitem{wardil_csf13}
\Name{Wardil L. \and da~Silva J. K.~L.} \REVIEW{Chaos, Solitons \& Fractals}{56}{2013}{160}.

\bibitem{maynard_82}
\Name{Maynard~Smith J.} \Book{Evolution and the Theory of Games} (Cambridge
  University Press, Cambridge, U.K.) 1982.

\bibitem{weibull_95}
\Name{Weibull J.~W.} \Book{Evolutionary Game Theory} (MIT Press, Cambridge, MA)
  1995.

\bibitem{hofbauer_98}
\Name{Hofbauer J. \and Sigmund K.} \Book{Evolutionary Games and Population
  Dynamics} (Cambridge University Press, Cambridge, U.K.) 1998.

\bibitem{mestertong_01}
\Name{Mesterton-Gibbons M.} \Book{An Introduction to Game-Theoretic Modelling} (AMS, Providence, RI) 2001.

\bibitem{nowak_06}
\Name{Nowak M.~A.} \Book{Evolutionary Dynamics} (Harvard University Press,
  Cambridge, MA) 2006.

\bibitem{cressman_igtr00}
\Name{Cressman R., Gaunersdorfer A. \and J.F. Wen J.~F.} \REVIEW{Int. Game Theory Rev.}{2}{2000}{67}.

\bibitem{hardin_g_s68}
\Name{Hardin G.} \REVIEW{Science}{162}{1968}{1243}.

\bibitem{fudenberg_e86}
\Name{Fudenberg D. \and Maskin E.} \REVIEW{Econometrica}{54}{1986}{533}.

\bibitem{nowak_n93}
\Name{Nowak M.~A. \and Sigmund K.} \REVIEW{Nature}{364}{1993}{56}.

\bibitem{santos_prl05}
\Name{Santos F.~C. \and Pacheco J.~M.} \REVIEW{Phys. Rev. Lett.}{95}{2005}{098104}.

\bibitem{wang_wx_pre08}
\Name{Wang W.-X., L{\"u} J., Chen G. \and Hui P.~M.} \REVIEW{Phys. Rev. E}{77}{2008}{046109}.

\bibitem{santos_pnas06}
\Name{Santos F.~C., Pacheco J.~M. \and Lenaerts T.} \REVIEW{Proc. Natl. Acad.
  Sci. USA}{103}{2006}{3490}.

\bibitem{fu_epjb07}
\Name{Fu F., Liu L.-H. \and Wang L.} \REVIEW{Eur. Phys. J. B}{56}{2007}{367}.

\bibitem{tanimoto_pre07}
\Name{Tanimoto J.} \REVIEW{Phys. Rev. E}{76}{2007}{021126}.

\bibitem{gomez-gardenes_prl07}
\Name{G{\'o}mez-Garde{\~n}es J., Campillo M., Flor{\'{\i}}a L.~M. \and Moreno
  Y.} \REVIEW{Phys. Rev. Lett.}{98}{2007}{108103}.

\bibitem{pacheco_ploscb09}
\Name{Pacheco J.~M., Pinheiro F.~L. \and Santos F.~C.} \REVIEW{PLoS Comput Biol}{5}{2009}{}.

\bibitem{fu_pre08b}
\Name{Fu F., Hauert C., Nowak M.~A. \and Wang L.} \REVIEW{Phys. Rev. E}{78}{2008}{026117}.

\bibitem{wang_wx_c11}
\Name{Wang W.-X., Lai Y.-C. \and Armbruster D.} \REVIEW{Chaos}{21}{2011}{033112}.

\bibitem{antonioni_pone11}
\Name{Antonioni A. \and Tomassini M.} \REVIEW{PLoS ONE}{6}{2011}{e25555}.

\bibitem{rong_pre10}
\Name{Rong Z., Wu Z.-X. \and Wang W.-X.} \REVIEW{Phys. Rev. E}{82}{2010}{026101}.

\bibitem{fu_pre09}
\Name{Fu F., Wu T. \and Wang L.} \REVIEW{Phys. Rev. E}{79}{2009}{036101}.

\bibitem{tanimoto_pre12}
\Name{Tanimoto J., Brede M. \and Yamauchi A.} \REVIEW{Phys. Rev. E}{85}{2012}{032101}.

\bibitem{press_pnas12}
\Name{Press W. \and Dyson F.} \REVIEW{Proc. Natl. Acad. Sci. USA}{109}{2012}{10409}.

\bibitem{hilbe_pnas13}
\Name{Hilbe C., Nowak M. \and Sigmund K.} \REVIEW{Proc. Natl. Acad. Sci. USA}{110}{2013}{6913}.

\bibitem{fu_jtb10}
\Name{Fu F., Nowak M.~A. \and Hauert C.} \REVIEW{J. Theor. Biol.}{266}{2010}{358}.

\bibitem{yang_hx_njp14}
\Name{Yang H.-X., Rong Z. \and Wang W.-X.} \REVIEW{New J. Phys.}{16}{2014}{013010}.

\bibitem{szolnoki_pre14}
\Name{Szolnoki A. \and Perc M.} \REVIEW{Phys. Rev. E}{89}{2014}{022804}.

\bibitem{santos_md_srep14}
\Name{Santos M., Dorogovtsev S.~N. \and Mendes J. F.~F.} \REVIEW{Sci. Rep.}{4}{2014}{4436}.

\bibitem{wu_js_pa14}
\Name{Wu J., Hou Y., Jiao L. \and Li H.} \REVIEW{Physica A}{412}{2014}{169}.

\bibitem{wang_j_pre10b}
\Name{Wang J., Fu F. \and Wang L.} \REVIEW{Phys. Rev. E} {82}{2010}{016102}.

\bibitem{hauser_jtb14}
\Name{Hauser O.~P., Traulsen A, \and Nowak M.~A.} \REVIEW{J. Theor. Biol.}{343}{2014}{178}.

\bibitem{perc_bs10}
\Name{Perc M. \and Szolnoki A.} \REVIEW{BioSystems}{99}{2010}{109}.

\bibitem{szabo_pr07}
\Name{Szab{\'o} G. \and F{\'a}th G.} \REVIEW{Phys. Rep.}{446}{2007}{97}.

\bibitem{roca_plr09}
\Name{Roca C.~P., Cuesta J.~A. \and S{\'a}nchez A.} \REVIEW{Phys. Life Rev.}{6}{2009}{208}.

\bibitem{schuster_jbp08}
\Name{Schuster S., Kreft J.-U., Schroeter A. \and Pfeiffer T.} \REVIEW{J. Biol.
  Phys.}{34}{2008}{1}.

\bibitem{santos_jtb12}
\Name{Santos F.~C., Pinheiro F., Lenaerts T. \and Pacheco J.~M.} \REVIEW{J.
  Theor. Biol.}{299}{2012}{88}.

\bibitem{perc_jrsi13}
\Name{Perc M., G{\'o}mez-Garde{\~n}es J., Szolnoki A., Flor{\'{\i}a} L.~M. \and Moreno Y.} \REVIEW{J. R. Soc. Interface}{10}{2013}{20120997}.

\bibitem{rand_tcs13}
\Name{Rand D.~A. \and Nowak M.~A.} \REVIEW{Trends in Cognitive Sciences}{17}{2013}{413}.

\bibitem{wu_t_pa14}
\Name{Wu T., Fu F., Dou P. \and Wang L.} \REVIEW{Physica A}{413}{2014}{86}.

\bibitem{van-doorn_jtb14}
\Name{van Doorn G.~S., Riebli T. \and Taborsky M.} \REVIEW{J. Theor. Biol.}{356}{2014}{1}.

\bibitem{barabasi_s99}
\Name{Barab{\'a}si A.-L. \and Albert R.} \REVIEW{Science}{286}{1999}{509}.

\bibitem{nowak_n92b}
\Name{Nowak M.~A. \and May R.~M.} \REVIEW{Nature}{359}{1992}{826}.

\end{thebibliography}
\end{document}